\documentstyle[12pt]{article}
\newcommand {\eq}{\begin{equation}}
\newcommand {\qe}{\end{equation}}
\newcommand {\cen}[1]{\begin{center} #1 \end{center}}

\begin{document}
\vspace*{-1.5in}
\vspace*{1.5in}
\cen{\bf \Huge A Parallel/Recursive Algorithm}
\cen {\bf \Large W. R. Gibbs}
\cen{New Mexico State University, Las Cruces NM, 88003}
\vspace*{1in}
\cen{Abstract}

An algorithm is discussed for converting a class of recursive processes to
a parallel system. It is argued that this algorithm can be superior to
certain methods currently found in the literature for an important subset
of problems. The cases of homogeneous and non-homogenous two term
recursion are treated. The basic cost factor of the algorithm over
non-parallel operations is 2 if only the final values of the sequence is
needed and 4 if all elements are required. In practice these factors can
be reduced considerably.  Applications to three problems (finding the
eigenvalues of a tri-diagonal matrix, the solution of a radial wave
equation and the solution of a tri-diagonal matrix) are discussed.

\newpage

\section{Introduction}

The solution of some problems require a greater number of operations for
an appropriate parallel algorithm than one that would be used in a
strictly serial calculation. A gain in speed can still be expected by
running them on a parallel machine but there is a cost factor since
one can not expect to achieve an increase in speed equal to the number of
processors when compared with the best non-parallel algorithm.

As an example, consider two term iteration.
\eq
x_{i+1}=a_ix_i+b_ix_{i-1} \ \ i=1,2,\dots N \label{one}
\qe

One method suggested in the literature\cite{modi} is to replace the steps
in the algorithm by matrix multiplication.  This algorithm requires extra
operations which will be discussed further at the end of section 2.1

Another possible algorithm (the one considered here) is based on the fact
that there are only two independent solutions to Eq. \ref{one}. The proper
linear combination of them to represent the actual solution can be
determined by the starting values.  In this paper the application of such
an algorithm for Eq. \ref{one}, as well as a similar one for the iteration
when there is an additional term, $c_i$, on the right hand side, is
discussed.

\section{General Description of the Algorithm}

\subsection{Homogeneous Case\label{homo}}

A recursion relation, such as Eq. \ref{one}, can be viewed as one long
sequence of values which leads from a beginning pair of values to the end.  
A desirable procedure for a parallel system would be to cut up this
sequence into separate strips (as many as there are processors) and let
each processor work through its part independently. For the first
processor there is no problem since the starting values are known there.
But the second processor (and the rest) will not have available their
starting values (the final values in the previous processor)  so this
procedure doesn't seem possible.  With a moderate expense, however, it can
be done. Since there are only two independent solutions of Eq. \ref{one}
we can construct two (arbitrary but independent) solutions, which will
provide basis functions, and combine them when the starting values for
each processor are known from the result of the previous one.  For
simplicity, consider the same algorithm running on all processors ignoring
the fact that it could be computed more efficiently on the first 
processor.

Let the total length of the recursion relation be $N+2$ (including the
first two starting values, hence the 2) with $M$ processors. Each
processor will be assigned a recursion of length $L=N/M$ which is supposed
to be integer and large. The work to be done by each processor will be
proportional to $M$ since it never has to calculate the first two values.
In order to see how such an algorithm works, let us analyze a (very
modest) system of 32 recursion steps to be calculated with 4 processors.

Each processor will do the recursion twice, once with starting values 0 and
1 and once with values 1 and 0. That is, each processor calculates the
two basis function starting with the first two values (1,0) and (0,1). It
uses the appropriate values of $a_i$ and $b_i$ for its position in the
global sequence, of course. For the first processor the basis functions 
start with
\eq
y_0^{10}=1;\ y_1^{10}=0;\ \ {\rm and}\ \ y_0^{01}=0;\ y_1^{01}=1.
\qe

Since any solution of the recursive formula can be written as a linear
combination of the two basis functions
\eq
x_i=\alpha y^{10}_i+\beta y^{01}_i,
\qe
we can see from the definition of the initial values in the first
processor that
\eq
x_0=\alpha;\ \ x_1=\beta,
\qe
where $x_0$\ and $x_1$\ are the starting values for Eq. \ref{one}.
We could find all of the values of the function in the first processor 
by calculating
\eq
x_i=x_0 y^{10}_i+x_1 y^{01}_i\ \ \ \ \ {\rm First\ Processor}
\ \ i=0,1,\dots ,8,9
\qe
but it is better not to do that immediately. Since each value is
independent, we may calculate only the last two values if we wish.  These
would be, in this simple case, $x_8$ and $x_9$.  Notice that these are the
starting values for the second processor.  Thus, for the second processor
since it started with $y_8^{10}=1;\ y_9^{10}=0$\ and $y_8^{01}=0;\
y_9^{01}=1$\ choosing the proper linear combination ($\alpha$ and $\beta$)
to give the true values of $x_8$ and $x_9$ (known from the first
processor) again we could calculate all of the values

\eq
x_i=x_8 y^{10}_i+x_9 y^{01}_i\ \ \ {\rm Second\ Processor}
\ \ i=8,9,\dots ,16,17.
\qe
Again, we need calculate only the last two ($x_{16}$ and $x_{17}$)
to get the starting values for the third processor. From these
we obtain the last values in the third processor $x_{24}$ and $x_{25}$
and the fourth processor $x_{32}$ and $x_{33}$. Thus, we have found the
last two values of the sequence with the evaluation (after the parallel
computations) of 8 equations. For 4 processors there will always be 8 
equations regardless of the length of iteration, $L$, within each processor.

Table 1 gives the operations explicitly for this small example.  
In this case each processor has only 8 iterations to do.  In a more
practical example numbers more like $10^6$ might be expected. The table
lists the initial conditions at the top followed by the iterations. The
generic variable, $y$, indicates both $y^{01}$ and $y^{10}$ are to be
calculated.

\begin{table}[htb]
$$
\begin{array}{|c|c|c|c|}
\hline
{\rm Processor\ 0}&{\rm Processor\ 1}&
{\rm Processor\ 2}&{\rm Processor\ 3}\\
\hline
y^{01}_0=0;\ y^{01}_1=1&y^{01}_8=0;\ y^{01}_9=1&
y^{01}_{16}=0;\ y^{01}_{17}=1&y^{01}_{24}=0;\ y^{01}_{25}=1\\
y^{10}_0=1;\ y^{10}_1=0&y^{10}_8=1;\ y^{10}_9=0&
y^{10}_{16}=1;\ y^{10}_{17}=0&y^{10}_{24}=1;\ y^{10}_{25}=0\\
\hline
y_2=a_1y_1+b_1y_0&y_{10}=a_{9}y_9+b_{9}y_8&
y_{18}=a_{17}y_{17}+b_{17}y_{16}&y_{26}=a_{25}y_{25}+b_{25}y_{24}\\
y_3=a_2y_2+b_2y_1&y_{11}=a_{10}y_{10}+b_{10}y_9&
y_{19}=a_{18}y_{18}+b_{18}y_{17}&y_{27}=a_{26}y_{26}+b_{26}y_{25}\\
\cdots&\cdots&\cdots&\cdots\\
\cdots&\cdots&\cdots&\cdots\\
\cdots&\cdots&\cdots&\cdots\\
y_7=a_6y_6+b_6y_5&y_{15}=a_{14}y_{14}+b_{14}y_{13}&
y_{23}=a_{22}y_{22}+b_{22}y_{21}&y_{31}=a_{30}y_{30}+b_{30}y_{29}\\
y_8=a_7y_7+b_7y_6&y_{16}=a_{15}y_{15}+b_{15}y_{14}&
y_{24}=a_{23}y_{23}+b_{23}y_{22}&y_{32}=a_{31}y_{31}+b_{31}y_{30}\\
y_9=a_8y_8+b_8y_7&y_{17}=a_{16}y_{16}+b_{16}y_{15}&
y_{25}=a_{24}y_{24}+b_{24}y_{23}&y_{33}=a_{32}y_{32}+b_{32}y_{31}\\
\hline
\end{array}
$$
\caption{Example of the algorithm for a small number of processors.}
\end{table}

After this work has been done the following sequential steps need to be
taken using only the last two values taken from each processor.

$$ x_8=x_0y_8^{10}+x_1y_8^{01};\ \  x_9=x_0y_9^{10}+x_1y_9^{01}$$
$$ x_{16}=x_8y_{16}^{10}+x_9y_{16}^{01};\ \  
x_{17}=x_8y_{17}^{10}+x_9y_{17}^{01}$$
$$ x_{24}=x_{16}y_{24}^{10}+x_{17}y_{24}^{01};\ \ 
 x_{25}=x_{16}y_{25}^{10}+x_{17}y_{25}^{01}$$
$$ x_{32}=x_{24}y_{32}^{10}+x_{25}y_{32}^{01};\ \ 
 x_{33}=x_{24}y_{33}^{10}+x_{25}y_{33}^{01}$$

The absolute indices have been used above on the basis functions. 
It is often more convenient to use a combination of the local index
and the processor number. The local index will be denoted by $\lambda$
and runs from 0 to $L+1$.

\begin{table}[htb]
$$
\begin{array}{rrrrrrrrrr}
 n&N=2^n&{\rm Sequential}&m=2&m=3&m=4&m=5&m=6&m=7&m=8\\
\hline
  7&    128&    384& 0.5625& 0.3750& 0.3750& 0.5625& 1.0312&1.0000&1.0000\\
  8&    256&    768& 0.5312& 0.3125& 0.2500& 0.3125& 0.5312&1.0156&1.0000\\
  9&    512&   1536& 0.5156& 0.2812& 0.1875& 0.1875& 0.2812&0.5156&1.0078\\
 10&   1024&   3072& 0.5078& 0.2656& 0.1562& 0.1250& 0.1562&0.2656&0.5078\\
 11&   2048&   6144& 0.5039& 0.2578& 0.1406& 0.0938& 0.0938&0.1406&0.2578\\
 12&   4096&  12288& 0.5020& 0.2539& 0.1328& 0.0781& 0.0625&0.0781&0.1328\\
 13&   8192&  24576& 0.5010& 0.2520& 0.1289& 0.0703& 0.0469&0.0469&0.0703\\
 14&  16384&  49152& 0.5005& 0.2510& 0.1270& 0.0664& 0.0391&0.0312&0.0391\\
 15&  32768&  98304& 0.5002& 0.2505& 0.1260& 0.0645& 0.0352&0.0234&0.0234\\
 16&  65536& 196608& 0.5001& 0.2502& 0.1255& 0.0635& 0.0332&0.0195&0.0156\\
 17& 131072& 393216& 0.5001& 0.2501& 0.1252& 0.0630& 0.0322&0.0176&0.0117\\
 18& 262144& 786432& 0.5000& 0.2501& 0.1251& 0.0627& 0.0317&0.0166&0.0098\\
 19& 524288&1572864& 0.5000& 0.2500& 0.1251& 0.0626& 0.0315&0.0161&0.0088\\
 20&1048576&3145728& 0.5000& 0.2500& 0.1250& 0.0626& 0.0314&0.0159&0.0083\\
\end{array}$$
\caption{Strip iteration algorithm. Column 3 gives the time to completion 
for straight iteration in one processor in units of the time required for
one floating point operation. Columns 4 through 10 give the ratio of the 
time to completion to the corresponding time for a purely sequential 
realization on a single processor (column 3). The number of iterations is 
$N=2^n$ with a number of processors $M=2^m$. This ``power of two'' 
representation is only for simplicity and is not needed for the algorithm.
}\label{strip}
\end{table}

In general, only two values in any processor need be computed to find the
starting values and hence the coefficients of the two basis functions for
the next processor.  So the serial overhead is only twice 3 floating point
operations per processor. Even if all of the values of the sequence are
needed, it is better to do this operation first, because the intermediate
values can then be found in parallel using the starting values obtained in
this way.

Repeating the above argument for the general case with $M$ processors 
labeled $\mu=0,1,2,\dots,M-1$ and $N+2$ total values of the indices of 
$x_i$, the resulting sequences are calculated in a (long) parallel 
calculation,
\eq
^{\mu}y^{10}_{\lambda}\ \ {\rm and}\ \ ^{\mu}y^{01}_{\lambda}
;\ \ \ \lambda=2,3,\dots ,L+1;\ \ [\mu=0,1,\dots ,M-1].
\qe
The square brackets indicate that the calculations for the different
values of $\mu$ are done in parallel. After this step, the equations
\eq
\left\{ 
\begin{array}{l}
x_{(\mu+1)L}=x_{\mu L}\ ^{\mu}y^{10}_L+x_{\mu L+1}\ ^{\mu}y_L^{01}\\
x_{(\mu+1)L+1}=x_{\mu L}\ ^{\mu}y^{10}_{L+1}+x_{\mu L+1}\ 
^{\mu}y_{L+1}^{01}\\
\end{array}\right\}_{\mu=0,1,\dots ,M-1}
\label{matching}\qe
are evaluated in a (short) sequential calculation. The number of 
equations is always twice the number of processors. The values of $x$ 
have been written with absolute indices but we may  
use a notation for individual processors. The pre-superscript $\mu$ 
as used above denotes results from a given processor $\mu$ so we could 
write equally valid representations of $x$ as
\eq
x_{\mu L+\lambda}=^{\mu}\!x_{\lambda}.
\qe
In matrix notation we may write Eqs. \ref{matching} as
\eq
\left(\begin{array}{c}^{\mu+1}x_0\\ ^{\mu+1}x_1\end{array}\right)=
\left(\begin{array}{cc}^{\mu}y^{10}_L&^{\mu}y^{01}_L\\
^{\mu}y^{10}_{L+1}&^{\mu}y^{01}_{L+1}\end{array}\right)
\left(\begin{array}{c}^{\mu}x_0\\ ^{\mu}x_1\end{array}\right)
_{\mu=0,1,\dots ,M-1}
\qe

Table \ref{strip} shows the ratios of completion times to the purely
sequential case to be expected with various recursion lengths and number
of processors. The increase in speed over scalar is $M/2$ for large $N$.
One sees that for moderate numbers of iterations the scaling efficiency
starts to fall of for a number of processors beyond 64.  Thus, for most
practical problems, the algorithm is expected to work best for relatively
large numbers of iterations and a modest number of processors.  This loss
comes, of course, because of the need to compute the matching relations
which requires a time proportional to the number of processors.  The
algorithm could be modified to spread this matching procedure over a
number of processors but this extention is beyond the scope of the present
work.

If only the end value of the sequence is needed one can stop at this point
(the ``short form'' of the algorithm).  This factor of two cost is not the
best that can be obtained if the values of $a_i$ and $b_i$ are being
calculated along with the iteration. The same values of these coefficients
are used in each iteration and, if the time for the calculation of the
coefficients is significant, the overhead to calculate two iterations
rather than one (as would happen if the calculation were not in parallel)
may be small.

The latency part of the communication time is proportional to $L=M/N$, so, 
for a fixed, moderate, number of processors and large $N$ it may be made 
very small. There are four words per processor to be sent in order to 
make the connection between segments.

At this point (if needed) one can proceed to calculate the entire sequence
of values in a second parallel calculation (the ``long form'' of the 
algorithm).  These will be given by
\eq
^{\mu}x_{\lambda}=x_{\mu L+\lambda}=x_{\mu L}\ ^{\mu}y^{10}_{\lambda}+x_{\mu 
L+1}\ ^{\mu}y_{\lambda}^{01}
;\ \ \ \lambda=2,3,\dots ,L+1;\ \ [\mu=0,1,\dots ,M-1]
\qe

These evaluations come at the cost of an additional $3L$ floating point
operations per processor, roughly a cost factor of 4 i.e., the speed
increase is $M/4$ compared to the pure sequential algorithm. This cost can
be reduced greatly in certain cases as we shall see later. It may be
useful to leave the strip functions (or even the strip basis functions) in
the processor where they were calculated.

We can now compare with the matrix algorithm mentioned in the introduction.
If we define
\eq y_i=
\left(\begin{array}{c}
x_{i+1}\\
x_i
\end{array}\right);
\ \ \ 
\alpha_i=\left(\begin{array}{cc}
a_i&b_i\\
1&0\end{array}\right),
\qe
then
\eq
y_i=
\left(\begin{array}{c}
x_{i+1}\\
x_i
\end{array}\right)= 
\left(\begin{array}{cc}
a_i&b_i\\
1&0\end{array}\right)
\left(\begin{array}{c}
x_i\\
x_{i-1}
\end{array}\right)=\alpha_iy_{i-1}
\qe
and the end member of the sequence is given by
\eq
\left(\begin{array}{c}
x_{N+1}\\
x_{N}
\end{array}\right)=y_N=
\alpha_N\alpha_{N-1}\alpha_{N-2}\cdots\alpha_1
\left(\begin{array}{c}
x_1\\    
x_{0}
\end{array}\right)
\qe

The multiplication of matrices can be done pairwise on different
processors.  The first multiplication of the simple matrices can be done
with two multiplications and one addition but this operation generates
full two by two matrices so that the second step in the pairwise reduction
of the multi-factor product is a complete matrix multiplication and
requires 8 multiplications and 4 additions. When compared with the 2
multiplications and 1 addition necessary to actually do the iteration, one
sees that there is a cost of a factor of 4 which is paid in this case.
Assuming 12 operations for all steps the maximum speed up with $M$
processors is $M/4$. The matrix algorithm gives only the end point of the
sequence (i.e. not the intermediate values) so is to be compared with
$M/2$ from the algorithm just presented. With a modest number of
processors, over half of the time of the matrix algorithm is spent in the
first set of multiplications so that considerable savings can be achieved
by considering the special case for the first operation.

This cost factor is not the only problem. The work done in each matrix
multiplication is not very much (12 floating point operations).  Hence,
communication must take place between processors very often so that
message passing time may dominate.

Another possible problem is that often the entire sequence of $x_i$
is needed. This algorithm simply doesn't give it.

\subsection{Inhomogeneous Recursion Relation \label{inhomo}}

For the inhomogeneous recursion

\eq 
x_{i+1}=a_ix_i+b_ix_{i-1}+c_i \label{inhomoeq} 
\qe
3 basis solutions are needed to provide a general representation. For
the third basis function we can take $z_i^{00}$, defined to have the 
starting conditions $z_{\mu L}^{00}=z_{\mu L+1}^{00}=0$. The general form 
of the solution is thus
\eq 
x_i=\alpha z_i^{10}+\beta z_i^{01}+\gamma z_i^{00}
\qe
where $z_i^{10}$, $z_i^{01}$, and $z_i^{00}$ all separately satisfy Eq. 
\ref{inhomoeq}. The requirement that this form satisfies the recursion 
relation is $\alpha+\beta+\gamma=1$.

Using this last expression to replace $\gamma$ we can write
\eq
x_i=\alpha (z_i^{10}-z_i^{00})+\beta(z_i^{01}-z_i^{00})+z_i^{00}.
\qe
It is easy to show that 
\eq
y_i^{10}\equiv z_i^{10}-z_i^{00}\ \ {\rm and}\ \   y_i^{01}\equiv 
z_i^{01}-z_i^{00}
\qe
satisfy the homogeneous equation (i.e. with $c_i=0$) with the same 
starting points as the $z_i^{10}$ and $z_i^{01}$, exactly as in the 
homogeneous case treated before. Thus, the general form can be written as
\eq
x_i=\alpha y_i^{10}+\beta y_i^{01}+z_i^{00} \label{gen3}
\qe
where $y_i^{10}$\ and $y_i^{01}$ can be calculated as in the previous 
section, i.e. without reference to $c_i$.

The procedure in this case is first to calculate in parallel
(either along with the $^{\mu}y_{\lambda}^{ij}$ or separately)
the (long) recursion
\eq
^{\mu}z^{00}_{\lambda+1}=a_{\mu L+\lambda}\ ^{\mu}\!z^{00}_{\lambda}+
b_{\mu L+\lambda}\  ^{\mu}\!z_{\lambda-1}^{00}+c_{\mu L+\lambda}
;\ \ \ \lambda=1,2,3,\dots ,L;\ \ [\mu=0,1,\dots ,M-1]\label{zcalc}.
\qe
Since both starting values of $^{\mu}z_{\lambda}^{00}$ are zero, the 
values of the coefficients of $y_i^{10}$\ and $y_i^{01}$ are again just 
the last values from the previous processor so that the equations
\eq
\left\{
\begin{array}{l}
x_{(\mu+1) L}=x_{\mu L}\ ^{\mu}y^{10}_L+x_{\mu L+1}\ ^{\mu}y_L^{01}
+^{\mu}\!\!z_L^{00}\\
x_{(\mu+1) L+1}=x_{\mu L}\ ^{\mu}y^{10}_{L+1}+x_{\mu L+1}\
^{\mu}y_{L+1}^{01}+^{\mu}\!\!z_{L+1}^{00}\\
\end{array}\right\}_{\mu=0,1,\dots ,M-2}
\qe
or in processor notation,
\eq
\left\{
\begin{array}{l}
^{(\mu+1)}x_0=^{\mu}\!x_0\ ^{\mu}y^{10}_L+^{\mu}\!x_1\ ^{\mu}y_L^{01}
+^{\mu}\!\!z_L^{00}\\
^{(\mu+1)}x_1=^{\mu}\!x_0\ ^{\mu}y^{10}_{L+1}+^{\mu}\!x_1\
^{\mu}y_{L+1}^{01}+^{\mu}\!\!z_{L+1}^{00}\\
\end{array}\right\}_{\mu=0,1,\dots ,M-2}
\qe
need to be evaluated in a (short) sequential calculation. 

If needed, the intermediate values in the recursive sequence can now be 
evaluated in parallel. These  will be given by
\eq
x_{\mu L+\lambda}=x_{\mu L}\ ^{\mu}y^{10}_{\lambda}+x_{\mu L+1}\ 
^{\mu}y_{\lambda}^{01}+^{\mu}\!\!z_{\lambda}^{00}
;\ \ \ \lambda=2,3,\dots ,L+1;\ \ [\mu=0,1,\dots ,M-1]
\qe
or
\eq
^{\mu}x_{\lambda}=^{\mu}\!\!x_0\ ^{\mu}y^{10}_{\lambda}+^{\mu}\!\!x_1\ 
^{\mu}y_{\lambda}^{01}+^{\mu}\!\!z_{\lambda}^{00}
;\ \ \ \lambda=2,3,\dots ,L+1;\ \ [\mu=0,1,\dots ,M-1]
\qe

If the recursion relation is needed for a large number of functions,
$c_i$, with the same $a_i$\ and $b_i$, then the basis functions $y^{10}$\
and $y^{01}$\ need be calculated only once.

 \section{Applications}

\subsection{Eigenvalues of a tri-diagonal matrix}

The process of finding eigenvalues of a real tri-diagonal matrix plays a
central role in the solution of the eigenvalue problem of more general
real symmetric matrices.  Commonly, algorithms are converted to a parallel
environment either by having each processor search for an eigenvalue
(method A) \cite{basermann} or finding all of the eigenvalues by means of
divide and conquer algorithms (method B) \cite{tisseur}. It is often
useful be able to pick out only a few of the eigenvalues (the
lowest ones) which is the case considered here.
 
For a symmetric tri-diagonal matrix 
\eq A= \left(\begin{array}{cccccccccc}
a_1&b_1&0&0&0&\dots\\ b_1&a_2&b_2&0&0&\dots\\ 0&b_2&a_3&b_3&0&\dots\\
&&&&&\dots\\ &&&&&\dots&b_{N-2}&a_{N-1}&b_{N-1}&0\\
&&&&&\dots&0&b_{N-1}&a_N&b_N\\ &&&&&\dots&0&0&b_N&a_{N+1}\\
\end{array}\right) 
\qe 
there exists a well-known solution for the eigenvalues, $\Lambda$, (see, 
for example, Refs.  \cite{gibbs,basermann}) based on Sturm sequences with 
bisection which allows the selection of eigenvalues.  With the definition
\eq
x_0=1;\ \ x_1=a_1-\Lambda \qe the recursion relation 
\eq
x_{i+1}=(a_{i+1}-\Lambda)x_i-b_i^2 x_{i-1} \label{eigen} 
\qe 
generates the determinant, $D(\Lambda)$, of $A-\Lambda I$ as the value of
$x_{N+1}$. The eigenvalues of $A$ can be found by locating the 
zeros of $D(\Lambda)$. Furthermore, the number of sign differences 
between successive members in the recursion sequence identifies the 
eigenvalues in order. For example, the lowest eigenvalue occurs at the 
transition from 0 to 1 sign differences in the sequence.  The desired 
transition (and hence eigenvalue) can be found with Newton's method of 
bisecting some $\Lambda_{min}$ and $\Lambda_{max}$ at each step. 

Implementing this recursion in a parallel fashion is straightforward using
the algorithm given in Section \ref{homo}. In order to calculate the
number of sign differences, the individual members of the sequence need to
be generated which requires the long form of the algorithm, thus seeming
to cost a factor of 4 compared to a non-parallelized version.  However, a
hybrid method makes the cost factor closer to 2 than 4. When the
difference in sign count has been reduced to unity between
$\Lambda_{min}$\ and $\Lambda_{max}$ it is known that a single eigenvalue
lies in this region and that it is the correct one. From this point on,
the method needs only the final value of the sequence (the determinant
itself) which requires only half the time.

Hence, the algorithm can be thought of as proceeding in two phases. In
the first phase the desired eigenvalue is isolated by finding two values
of the estimated eigenvalue with only a single zero of the determinant
between them. After that, in the second phase, only the value of the 
determinant is needed and with those values one can estimate a new trial 
value more efficiently than a simple bisection by using
\eq
\Lambda=\frac{\Lambda_{min}|D(\Lambda_{max})|+\Lambda_{max}|D(\Lambda_{min})|}
{|D(\Lambda_{max})|+|D(\Lambda_{min})|}.
\qe
One must be careful of the convergence since it may come about with the
trial eigenvalue approaching one of the limit eigenvalues without the
two limits approaching each other. This improvement in efficiency is 
available to either the parallel or non-parallel version and tends to make 
the first phase dominant in time consumed.

A common method of implementing this general algorithm on parallel
computers is to simply give each processor an eigenvalue to find (method A
above). In this case each eigenvalue is obtained in a purely sequential
fashion but the values arrive in a parallel manner.  In the present
algorithm each eigenvalue is calculated in a parallel manner and the
values arrive one after the other. One improvement which is not generally
available to method A is due to the availability of useful information
after the first and subsequent eigenvalues are found in the first phase.
It is only necessary to keep a table of the tested values of $\Lambda$ vs.
the corresponding number of sign differences.  When the next
eigenvalue is to be found, the table can be searched for the 
closest starting values.  This table grows as the eigenvalues are found.
If each processor is finding an eigenvalue starting from the outer bounds
of the eigenvalue sequence this advantage is not available.

An important consideration in this algorithm (parallel or non-parallel) is
the growing of the values with each step so that overflow occurs.  One can
solve that problem by performing a renormalization at regular intervals.  
In the non-parallel version, when the value of $x_{i+1}$ is observed to
exceed some predefined value then both $x_{i+1}$ and $x_{i}$ are
multiplied by an appropriate constant reducing factor. This does not
change the number of relative sign differences nor the sign of the last
value.  For the parallel version both basis functions can be renormalized
in the same way (both must be done at the same time) and the recursion is
not destroyed since it is only the relative values of the basis functions
which determines the number of sign differences and the sign of the final
value. In the version tested here a check for renormalization was 
made every 16 steps.

The algorithm was calculated for the matrix defined by 
\eq 
a_i=0,\ i=1,2,\dots N+1;\ \  b_i^2=i(N+1-i),\ \ i=1,2,\dots N
\qe
with N even. The eigenvalues are known to be the even integers from --N to
N as presented in Refs. \cite{basermann,gregory}. In addition to the
renormalization mentioned above, the entire system was renormalized such
that the smallest $b_i^2$ was unity. Calculations were made for matrices
of size up to $N=10,240,000$ finding eigenvalues to an accuracy of 1 part
in $10^{11}$. The method was tested on a Beowulf cluster with a 100 Mbit
Ethernet (using MPICH\cite{mpich,mpicha}) and the scaling efficiency
[defined as $T_1/(MT_M)$ where $T_M$ is the time for execution on $M$
processors] exceeded 0.96 up through 4 processors.

The ``cost'' of the method was calculated by comparing one-processor
versions of the present algorithm with a simple sequential calculation but
with the present algorithm taking advantage of the information contained
in the table of the number of sign differences vs. trial $\Lambda$.  For a
single eigenvalue the present method takes about 2.5 times longer than
method A. For 5 eigenvalues it still takes about 20\% longer. For 10
eigenvalues it is 0.85 as long, for 20 it is 0.72 as long and for 40 it
requires 0.64 of the time for method A.

Procedure A also may suffer from an incommensurability with the number of
processors. If one wishes 8 eigenvalues with 64 processors then only 1/8
of the capacity of the machine is being used.  The calculation of the
relative efficiency of the methods can be quite complicated, however. For
example, if one wishes 16 eigenvalues with 8 processors then two passes
will be made with method A and in the second pass tables of values
accumulated in the first pass can be used. Incorporating this information
could well make method A faster.

\subsection{Solving wave equations}

Wave equations (the Schr\"odinger equation is considered here) 
can be solved, after an expansion in Legendre polynomials,
by means of one-dimensional second order differential equations. An 
accurate solution can be obtained with Noumerov's method 
where the iteration equations for the reduced wave function are given by

\eq
\psi_{i+1}=\frac{2\psi_i-\psi_{i-1}-\frac{h^2}{12}\left[ 10w_i\psi_i
+w_{i-1}\psi_{i-1}\right]}{1+\frac{h^2}{12}w_{i+1}},
\qe
where

\eq
w(r)=k^2-\frac{2m}{\hbar^2}V(r)-\frac{\ell(\ell+1)} {r^2},
\qe
and $h$ is the spacing in the radial variable $r$.

This form is converted readily into that considered in Section \ref{homo}
with much of the work of the computation of $a_i$ and $b_i$ being done
before the iteration.  Keeping $k^2$ as a free parameter and precomputing
one vector \{$1-h^2[\frac{2m}{\hbar^2}V(r)+\frac{\ell(\ell+1)}{r^2}]\}$
the calculation of $a_i$ and $b_i$ requires seven floating point
operations. Given that each iteration requires an additional three
operations, we might expect that the parallel algorithm involves a cost of
an increase from ten to thirteen operations or about 30\% over the purely
sequential one.

In the present case the Schr\"{o}dinger equation is solved for a bound
state. The solution is started at the origin with zero for the first point
and an arbitrary value for the second point and a search is made for a
value of the energy (expressed here as $k^2=2mE/\hbar^2$) that causes the
wave function at some large value of $r$ to be zero.

The problem was treated with a near maximum precomputation
of the values of $a_i$ and $b_i$. Clearly, if one chooses not to compute
as much as possible in advance, and hence to spend a larger amount of time
in the computation of the coefficients, then the addition of a second
iteration (the expense of this algorithm) would make less difference.  
Thus, a true test of the algorithm relies on a realistic degree of
precomputation.

The calculation was coded and tested for $\ell=2$. The number of steps
taken was $25.2 \times 10^6$.  Because of the boundary conditions of the
problem (reduced wave function zero at the origin) the calculation of the
basis function $y^{10}$ is not needed in the first processor. By comparing
a one-processor calculation with it computed or not it was found that the
cost factor of the algorithm was 1.31, in agreement with the 30\% increase
estimated above.  The method was tested on the Beowulf cluster and no
decrease in scaling efficiency was seen through 16 processors.

\subsection{Tri-diagonal Matrix Solution}

Consider the recursive solution to a tri-diagonal matrix (size $N+2$), at
first without any parallelism.

\eq \left(
\begin{array}{cccccccccc}
a_0&e_0&0&0&0&\dots\\
b_1&a_1&e_1&0&0&\dots\\
0&b_2&a_2&e_2&0&\dots\\
&&&&&\dots\\
&&&&&\dots\\
&&&&&\dots\\
&&&&&\dots&b_{N-1}&a_{N-1}&e_{N-1}&0\\
&&&&&\dots&0&b_N&a_N&e_N\\
&&&&&\dots&0&0&b_{N+1}&a_{N+1}\\
\end{array}\right)
\left(\begin{array}{c}
x_0\\
x_1\\
x_2\\
\dots\\
\dots\\
\dots\\
x_{N-1}\\
x_{N}\\
x_{N+1}\\
\end{array}\right)
=\left(
\begin{array}{c}
c_0\\
c_1\\
c_2\\
\dots\\ 
\dots\\
\dots\\
c_{N-1}\\
 c_N\\ c_{N+1}\\
\end{array}\right)
\qe

If any $e_i$ (let the first occurrence be at $i=k$) is zero then the
system can be reduced to two subsystems. To see this observe that the
first equation alone consists of one equation in two unknowns, the first
two equations correspond to two equations in three unknowns, etc.  If
$e_k=0$ then adding that equation to the system introduces no new unknown
so the system of the first $k+1$ equations can be solved alone giving the
value (among others) of $x_k$. In the remaining equations the $k^{th}$
column can be taken to the right hand side so that they can be solved.  
In this case the system is separable. For a symmetric system, if $e_k=0$
then $b_{k+1}=0$ also and the two blocks are completely decoupled.  Here
it is assumed that this is NOT the case so that NO $e_i=0$. Thus, we can
divide all equations by $e_i$ or, equivalently, we can set $e_i=1$ in the
system we wish to consider.  Of course, the conversion of a general system
to this form entails the cost of one inverse and two multiplications per
equation on the left hand but needs to be done only once in the case of a
number of different right hand sides ($N+1$ more multiplications are
necessary for each right hand side).

For these reasons, the following system is
considered.

\eq \left(
\begin{array}{cccccccccc}
a_0&1&0&0&0&\dots\\
b_1&a_1&1&0&0&\dots\\
0&b_2&a_2&1&0&\dots\\
&&&&&\dots\\
&&&&&\dots\\
&&&&&\dots\\
&&&&&\dots&b_{N-1}&a_{N-1}&1&0\\
&&&&&\dots&0&b_N&a_N&1\\
&&&&&\dots&0&0&b_{N+1}&a_{N+1}\\
\end{array}\right)
\left(\begin{array}{c}
x_0\\
x_1\\
x_2\\
\dots\\
\dots\\
\dots\\
x_{N-1}\\
x_{N}\\
x_{N+1}\\
\end{array}\right)
=\left(
\begin{array}{c}
c_0\\
c_1\\
c_2\\
\dots\\ 
\dots\\
\dots\\
c_{N-1}\\
 c_N\\ c_{N+1}\\
\end{array}\right)
\qe

Starting from the second row the equations can be expressed as the 
recursion relation
\eq
x_{i+1}=-a_ix_i-b_ix_{i-1}+c_{i};\ i=1,2,3, \dots ,N-1,N,
\label{recurm}\qe
where neither the first or last equations have been used.

The three basis solutions discussed in Section \ref{inhomo} (called here
$f_i^{10}$, $f_i^{01}$ and $g_i^{00}$) can be used to provide ``global''
basis functions (global in the sense that they represent the full
recursion sequence to be distinguished from the strip functions to be
discussed shortly) to express the solution. The first two basis solutions
do not involve $c_i$ and need only be calculated once for many right hand
sides.  Thus, the solution separates into two parts, somewhat similar to
the common factorization and back substitution methods.  Once we have the
basis solutions we can apply the conditions implied by the first and last
equation to determine the coefficients $\alpha$ and $\beta$ in Eq.  
\ref{gen3}. For the first equation we have

\eq
a_0x_0+x_1=a_0(\alpha f_0^{10} +\beta f_0^{01} +g_0^{00})
+\alpha f_1^{10}+\beta f_1^{01}+g_1^{00}=c_0
\label{31}\qe
or
\eq
a_0\alpha+\beta=c_0 \label{32}.
\qe
From the last equation we have
\eq a_{N+1}x_{N+1}+b_{N+1}x_N=
a_{N+1}(\alpha f_{N+1}^{10}+\beta f_{n+1}^{01}+g_{N+1}^{00})
+b_{N+1}(\alpha f_{N}^{10}+\beta f_{N}^{01}+g_{N}^{00})=c_{N+1}
\qe
or
\eq
\alpha(a_{N+1}f_{N+1}^{10}+b_{N+1}f_N^{10})+
\beta(a_{N+1}f_{N+1}^{01}+b_{N+1}f_N^{01})=c_{N+1}-a_{N+1}g_{N+1}^{00}
-b_{N+1}g_N^{00}
\label{34}\qe

From these two equations we obtain $\alpha$ and $\beta$ and all values
of $x_i$ can be obtained from Eq. \ref{gen3}.  

As an alternative to Eq. \ref{34} one can iterate one further step with
Eq. \ref{recurm} to obtain $f^{10}_{N+2},\ f^{01}_{N+2}\ {\rm and}\
g^{00}_{N+2}$\ and use the condition that $x_{N+2}=0$\ to find

\eq
\alpha f^{10}_{N+2}+\beta f^{01}_{N+2}=-g^{00}_{N+2}.
\qe

Returning to the parallel considerations, we can express the 
global basis functions in terms of the strip basis solutions
in each processor, obtain the three global functions that 
were used in the above algorithm and then calculate the solution. 
However, it is much more efficient to combine the two operations.

First write the global recursion basis functions in
terms of the strip basis functions.
\eq
f_{\mu L+\lambda}^{10}=\ ^{\mu}\!\alpha^{10}\ ^{\mu}y_{\lambda}^{10}
+^{\mu}\!\!\beta^{10}\ ^{\mu}y_{\lambda}^{01}
\label{35}\qe
\eq
f_{\mu L+\lambda}^{01}=\ ^{\mu}\!\alpha^{01}\ ^{\mu}y_{\lambda}^{10}
+^{\mu}\!\!\beta^{01}\ ^{\mu}y_{\lambda}^{01}
\label{36}\qe
\eq
g_{\mu L+\lambda}^{00}=\ ^{\mu}\!\alpha^{00}\ ^{\mu}y_{\lambda}^{10}
+^{\mu}\!\!\beta^{00}\ ^{\mu}y_{\lambda}^{01}+^{\mu}\!\!z_{\lambda}^{00}
\label{37}\qe
where the $^{\mu}\alpha$ and $^{\mu}\beta$ are to be obtained from the
matching conditions for $\mu=1,2,\dots, M-1$.
\eq
^{\mu}\alpha^{10}=\ ^{\mu-1}\alpha^{10}\ ^{\mu-1}y_L^{10}+
^{\mu-1}\!\beta^{10}\ ^{\mu-1}y_L^{01}
\qe
\eq
^{\mu}\beta^{10}=\ ^{\mu-1}\alpha^{10}\ ^{\mu-1}y_{L+1}^{10}+
^{\mu-1}\!\beta^{10}\ ^{\mu-1}y_{L+1}^{01}
\qe
\eq
^{\mu}\alpha^{01}=\ ^{\mu-1}\alpha^{01}\ ^{\mu-1}y_L^{10}+
^{\mu-1}\!\beta^{01}\ ^{\mu-1}y_L^{01}
\qe
\eq
^{\mu}\beta^{01}=\ ^{\mu-1}\alpha^{01}\ ^{\mu-1}y_{L+1}^{10}+
^{\mu-1}\!\beta^{01}\ ^{\mu-1}y_{L+1}^{01}
\qe
\eq
^{\mu}\alpha^{00}=\ ^{\mu-1}\alpha^{00}\ ^{\mu-1}y_L^{10}+
^{\mu-1}\!\beta^{00}\ ^{\mu-1}y_L^{01}+^{\mu-1}\!z_L^{00}
\qe
\eq
^{\mu}\beta^{00}=\ ^{\mu-1}\alpha^{00}\ ^{\mu-1}y_{L+1}^{10}+
^{\mu-1}\!\beta^{00}\ ^{\mu-1}y_{L+1}^{01}+^{\mu-1}\!z_{L+1}^{00}
\qe
with the starting values
\eq
^0\alpha^{10}=1;\ ^0\alpha^{01}=0;\ ^0\beta^{10}=0;\ ^0\beta^{01}=1;\ 
^0\alpha^{00}=\ ^0\!\beta^{00}=0.
\qe 
Using the last two values of the global basis functions calculated from 
Eq. \ref{35}-\ref{37} we can solve for the global $\alpha$ and $\beta$
(from Eq. \ref{32} and \ref{34}) to write

\eq
x_{\mu L+\lambda}=
^{\mu}\!\alpha\ ^{\mu}y^{10}_{\lambda}+^{\mu}\!\beta\ 
^{\mu}y^{01}_{\lambda}+^{\mu}\!z^{00}_{\lambda};\ \ 
\lambda=2,3,\dots,L+1\ \ [\mu=0,1,\dots,M-1]\label{finish}
\qe
where the coefficients are given by
\eq
^{\mu}\alpha=\alpha\ ^{\mu}\alpha^{10}+\beta\ ^{\mu}\alpha^{01}
+^{\mu}\!\alpha^{00}
\qe
\eq
^{\mu}\beta=\alpha\ ^{\mu}\beta^{10}+\beta\ ^{\mu}\beta^{01}
+^{\mu}\!\beta^{00}
\qe

It is common to compare the relative speed of any algorithm for
solving tri-diagonal matrices with Gaussian Elimination (GE) which is 
relatively efficient. For this case GE becomes, first for the LU
reduction
$$ d_0=1/a_0$$
\eq
g_i=b_id_{i-1};\ d_i=1/(a_i-g_i)\ \ \ i=1,2,\dots, N+1,
\qe
followed by the two back substitutions
\eq
z^g_0=c_0;\ z^g_i=c_i-g_iz^g_{i-1};\ \ i=1,2,\dots,N+1,
\label{bs1}
\qe
and
\eq
x^g_{N+1}=z^g_{N+1};\ x^g_i=(z^g_i-x^g_{i+1})d_i;\ \ i=N,N-1,\dots, 0.
\label{bs2}
\qe

If we assume that the equations are being solved for many right hand
sides, then we should compare the time for the solutions of the $z^{00}$
equations and the calculation of the vector (Eqs. \ref{zcalc} and
\ref{finish}) with the work of the two back substitutions in GE (Eqs.  
\ref{bs1} and \ref{bs2}). A first estimation can be made for the relative
speed by counting the number of floating point operations per step (4 for
GE and 8 for the parallel algorithm) to get a cost factor of 2. This is
only a very crude estimate since the form of the equations is different.
For example, Eq. \ref{finish} requires only the broadcast of a scalar
instead of vector multiplication.  Optimization or not of the G77 compiler
was observed to make a large difference also.  With no optimizing GE does
better than this estimate giving a cost factor of 2.6.  However, with
optimization, the efficiency of the parallel method is improved more than
GE to result in a cost factor of 1.4.

To save on message passing for the resultant vector, one may want to use 
the strips in the processor in which they were formed. In some cases
it may be more efficient never to construct the vectors at all. As an
example of such a case, consider the problem of solving the set of 
equations for a large number of different right hand side vectors which 
are a function of some parameter, $\eta$, hence, $c_i(\eta)$. Suppose also
that we wish the sum (an integral perhaps) of some weighting function over
the solution
\eq
S(\eta)=\sum_{i=0}^{N+1} w_ix_i(\eta)
\qe
as a function of $\eta$. The sum can be distributed among the strip basis
functions in the processors via Eq. \ref{finish}.  The $y$ basis function
integrals need be calculated only once.  The $z^{00}$ integral can be
calculated as this basis function is generated.  Only the strip {\it
integrals} need to be sent to the master processor to be combined with the
coefficients $^{\mu}\alpha$\ and $^{\mu}\beta$. The calculation of the
solution (Eq. \ref{finish}) is not needed. In this special case, a count
of the number of floating point operations estimates the speed to be the
same as GE (in the limit of large $N$ and large number of values of
$\eta$). In one-processor tests, because of the simplicity of the
equations mentioned above, the strip algorithm was found to run somewhat
faster than GE.

Large systems ($720,720\times k$ with $k=10,\ 20,\ 40,\ 80,\ {\rm and}\
160$) with 100 values of $\eta$ were tested on the 16 processor Beowulf
cluster. Essentially no degradation of performance was seen with all
scaling efficiencies $\geq 0.99$.  The largest system tested
($N=115,315,200$) could only be run by spreading the solution basis
vectors over 13 processors.

A common algorithm discussed in the literature is the parallel cyclic
reduction of a matrix\cite{nguyen}.  The basic cost of this algorithm
has been reported to be a factor of 4 \cite{hajj,reale}.  It requires
frequent exchange of information and is not very efficient for multiple
right hand sides. The ``divide and conquer'' method \cite{nguyen} is
also inefficient for multiple driving terms.  Hence, the technique
presented here would seem to offer an attractive alternative to these
methods in some cases.

The restriction to $e_i\ne 0$ may prove to be inconvenient in some
cases or the division may lead to large errors.  Tests with $e_i\equiv 1$
showed that the stability of the method was as good or better than GE. 

\section{Discussion}

These algorithms may also be useful on vector machines. For a processor
with a 64 word vector register, for the case of the homogeneous recursion
relation, the total length can be broken into 32 strips with each pair of
words in the vector register acting as a processor. Thus, the iteration
might take place as

\eq
\left(\begin{array}{c}
y^{10}_{i+1}\\
y^{01}_{i+1}\\
y^{10}_{L+i+1}\\
y^{01}_{L+i+1}\\
\dots\\
\dots\\
y^{10}_{\mu L+i+1}\\
y^{01}_{\mu L+i+1}\\
\dots\\
\dots\\
y^{10}_{31L+i+1}\\
y^{01}_{31L+i+1}\\
\end{array}\right)
=
\left(\begin{array}{c}
a_i\\
a_i\\
a_{L+i}\\
a_{L+i}\\
\dots\\
\dots\\
a_{\mu L+i}\\
a_{\mu L+i}\\
\dots\\
\dots\\
a_{31 L+i}\\
a_{31L+i}\\
\end{array}\right)
\times
\left(\begin{array}{c}
y^{10}_{i}\\
y^{01}_{i}\\
y^{10}_{L+i}\\
y^{01}_{L+i}\\
\dots\\
\dots\\
y^{10}_{\mu L+i}\\
y^{01}_{\mu L+i}\\
\dots\\
\dots\\
y^{10}_{31L+i}\\
y^{01}_{31L+i}\\
\end{array}\right)
+
\left(\begin{array}{c}
b_i\\
b_i\\
b_{L+i}\\
b_{L+i}\\
\dots\\
\dots\\
b_{\mu L+i}\\
b_{\mu L+i}\\
\dots\\
\dots\\
b_{31 L+i}\\
b_{31L+i}\\
\end{array}\right)
\times
\left(\begin{array}{c}
y^{10}_{i-1}\\
y^{01}_{i-1}\\
y^{10}_{L+i-1}\\
y^{01}_{L+i-1}\\
\dots\\
\dots\\
y^{10}_{\mu L+i-1}\\
y^{01}_{\mu L+i-1}\\
\dots\\
\dots\\
y^{10}_{31L+i-1}\\
y^{01}_{31L+i-1}\\
\end{array}\right)
\qe
for $i=1,2,\dots L$.

\vspace*{.3in}

The method can be generalized for a larger number of terms in the iteration 
(leading to larger width in banded matrices, for example).

There are clearly some limitations to the application of the algorithm.  
The conversion to a parallel system does not work for recursions non
linear in $x_i$ so most classical mechanics calculations are not possible
with it.

While the problems treated are different, this method appears to have some 
overlap with the Domain Decomposition techniques for the solution of
elliptic differential equations \cite{roache}.

I wish to thank Slava Solomatov for discussions and Alexei Vezolainen for
help with one of Beowulf clusters in the Dept. of Physics.  This work was
supported by the National Science Foundation under contract PHY-0099729.

 \end{document}